\documentclass[prl,nofootinbib,floatfix,twocolumn,showpacs,superscriptaddress]{revtex4}
\usepackage[dvips]{pstcol}
\usepackage{graphicx,amsmath}
\usepackage{dcolumn}
\usepackage{bm}
\usepackage{rotating}
\usepackage{amsmath,amssymb,amsfonts}
\usepackage{color}	

\def\hermes{{\sc Hermes~}}

\def\jlab{{\sc JLab~}}
\def\desy{{\sc Desy~}}
\def\hera{{\sc Hera~}}
\def\cornell{{\sc Cornell~}}
\def\cea{{\sc Cea~}}

\def\jetset{{\sc Jetset~}}
\def\pythia6{{\sc Pythia6~}}
\def\pythia{{\sc Pythia~}}
\def\radgen{{\sc Radgen~}}
\def\geant{{\sc Geant~}}
\newcommand{\reac}{$e p\to e n \pi^+ ~$}
\newcommand{\reacX}{$e p\to e \pi^+ X ~$}

\begin{document}

\title{Cross sections for hard exclusive electroproduction of $\pi^{+}$ mesons on a hydrogen target}

\def\groupargonne{\affiliation{Physics Division, Argonne National Laboratory, Argonne, Illinois 60439-4843, USA}}
\def\groupbari{\affiliation{Istituto Nazionale di Fisica Nucleare, Sezione di Bari, 70124 Bari, Italy}}
\def\groupbeijing{\affiliation{School of Physics, Peking University, Beijing 100871, China}}
\def\groupcolorado{\affiliation{Nuclear Physics Laboratory, University of Colorado, Boulder, Colorado 80309-0390, USA}}
\def\groupdesy{\affiliation{DESY, 22603 Hamburg, Germany}}
\def\groupzeuthen{\affiliation{DESY, 15738 Zeuthen, Germany}}
\def\groupdubna{\affiliation{Joint Institute for Nuclear Research, 141980 Dubna, Russia}}
\def\grouperlangen{\affiliation{Physikalisches Institut, Universit\"at Erlangen-N\"urnberg, 91058 Erlangen, Germany}}
\def\groupferrara{\affiliation{Istituto Nazionale di Fisica Nucleare, Sezione di Ferrara and Dipartimento di Fisica, Universit\`a di Ferrara, 44100 Ferrara, Italy}}
\def\groupfrascati{\affiliation{Istituto Nazionale di Fisica Nucleare, Laboratori Nazionali di Frascati, 00044 Frascati, Italy}}
\def\groupgent{\affiliation{Department of Subatomic and Radiation Physics, University of Gent, 9000 Gent, Belgium}}
\def\groupgiessen{\affiliation{Physikalisches Institut, Universit\"at Gie{\ss}en, 35392 Gie{\ss}en, Germany}}
\def\groupglasgow{\affiliation{Department of Physics and Astronomy, University of Glasgow, Glasgow G12 8QQ, United Kingdom}}
\def\groupillinois{\affiliation{Department of Physics, University of Illinois, Urbana, Illinois 61801-3080, USA}}
\def\groupmichigan{\affiliation{Randall Laboratory of Physics, University of Michigan, Ann Arbor, Michigan 48109-1040, USA }}
\def\groupnikhef{\affiliation{National Institute for Subatomic Physics (Nikhef), 1009 DB Amsterdam, The Netherlands}}
\def\groupstpetersburg{\affiliation{Petersburg Nuclear Physics Institute, St. Petersburg, Gatchina, 188350 Russia}}
\def\groupprotvino{\affiliation{Institute for High Energy Physics, Protvino, Moscow region, 142281 Russia}}
\def\groupregensburg{\affiliation{Institut f\"ur Theoretische Physik, Universit\"at Regensburg, 93040 Regensburg, Germany}}
\def\grouprome{\affiliation{Istituto Nazionale di Fisica Nucleare, Sezione Roma 1, Gruppo Sanit\`a and Physics Laboratory, Istituto Superiore di Sanit\`a, 00161 Roma, Italy}}
\def\grouptriumf{\affiliation{TRIUMF, Vancouver, British Columbia V6T 2A3, Canada}}
\def\grouptokyo{\affiliation{Department of Physics, Tokyo Institute of Technology, Tokyo 152, Japan}}
\def\groupamsterdam{\affiliation{Department of Physics and Astronomy, Vrije Universiteit, 1081 HV Amsterdam, The Netherlands}}
\def\groupwarsaw{\affiliation{Andrzej Soltan Institute for Nuclear Studies, 00-689 Warsaw, Poland}}
\def\groupyerevan{\affiliation{Yerevan Physics Institute, 375036 Yerevan, Armenia}}
\def\groupnone{\noaffiliation}

\groupargonne
\groupbari
\groupbeijing
\groupcolorado
\groupdesy
\groupzeuthen
\groupdubna
\grouperlangen
\groupferrara
\groupfrascati
\groupgent
\groupgiessen
\groupglasgow
\groupillinois
\groupmichigan
\groupnikhef
\groupstpetersburg
\groupprotvino
\groupregensburg
\grouprome
\grouptriumf
\grouptokyo
\groupamsterdam
\groupwarsaw
\groupyerevan

\author{A.~Airapetian}  \groupmichigan
\author{N.~Akopov}  \groupyerevan
\author{Z.~Akopov}  \groupyerevan
\author{E.C.~Aschenauer}  \groupzeuthen
\author{W.~Augustyniak}  \groupwarsaw
\author{A.~Avetissian}  \groupyerevan
\author{E.~Avetissian}  \groupfrascati
\author{L.~Barion}  \groupferrara
\author{S.~Belostotski}  \groupstpetersburg
\author{N.~Bianchi}  \groupfrascati
% ...

\collaboration{The HERMES Collaboration} \noaffiliation

\date{\today}

\begin{abstract}
The exclusive electroproduction of $\pi^+$ mesons was studied 
with the \hermes spectrometer at the \desy laboratory 
by scattering 27.6 GeV positron and electron beams off an internal hydrogen 
gas target. The virtual-photon cross sections were measured as a function of the 
Mandelstam variable $t$ and the squared four momentum $-Q^2$ of the exchanged virtual 
photon. A model calculation based on 
Generalized Parton Distributions is in fair agreement with the data at low values of $|t|$ 
if power corrections are included. A model calculation based on the 
Regge formalism gives a good description of the magnitude and the $t$ and $Q^2$ dependences 
of the cross section.
\end{abstract}


\maketitle

\begin{figure}[t]
\begin{center}
\includegraphics[width=0.45 \textwidth]{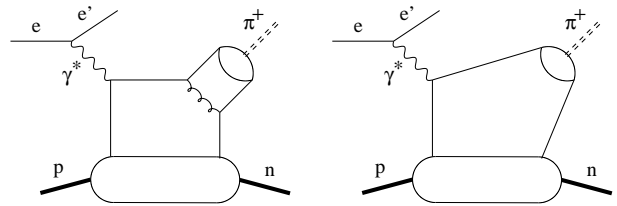}
\caption[]{
Leading-order (left panel) and soft-overlap (right panel) diagram for 
the exclusive $\pi^+$ electroproduction amplitude.
}
\label{fig0}
\end{center}
\end{figure}

The interest in hard exclusive processes has grown since a Quantum Chromodynamics (QCD) factorization 
theorem was proven for the hard electroproduction of mesons by longitudinal 
photons~\cite{Collins}. It was shown that at leading order in the strong-coupling constant $\alpha_S$ and in the fine-structure constant $\alpha$ the amplitude 
for such reactions can be factorized into three parts (see left panel of Fig.~\ref{fig0}): a hard lepton-scattering part, 
which can be calculated in Quantum Electrodynamics and in perturbative QCD (pQCD), and two 
soft parts that parametrize the structure of the target nucleon by Generalized Parton Distributions 
(GPDs)~\cite{Muller,Rady,Ji} and the structure of the produced meson by a distribution amplitude. 
The GPDs offer the 
possibility to reveal a three-dimensional representation of 
the structure of hadrons at the partonic level, correlating the longitudinal momentum fraction to transverse spatial coordinates~\cite{Burkardt,Diehl2002,Ralston,Belitsky2002,Burkardt2003}. For recent theoretical reviews, see Refs.~\cite{Goeke,Diehl,Belitsky}.

The amplitude for exclusive electroproduction of mesons with specific quantum numbers is described by a particular combination of GPDs. At leading twist, exclusive vector-meson production is sensitive to only unpolarized GPDs ($H$ and $E$), while pseudoscalar-meson production is sensitive to polarized GPDs ($\widetilde H$ and $\widetilde E$) without the need for a polarized target or beam.
Moreover, for $\pi^+$ production, the pseudoscalar contribution involving $\widetilde E$ dominates at small momentum transfer to the target as it contains the $t-$channel pion-pole contribution. The complete amplitude of that contribution to $\pi^+$ production also contains the pion charge form factor~\cite{Frazer,Sullivan}. 

At moderate virtuality of the exchanged photon, next-to-leading-order (NLO) 
in $\alpha_S$ and 
higher-twist corrections to the pQCD leading-order 
amplitude can contribute. Two different types of higher-twist corrections,
 jointly denoted by the term ``power corrections'', have been estimated in the case of $\pi^+$ 
production~\cite{VGG}: one arising from the intrinsic 
transverse momentum of the partons, and the other resulting from the soft-overlap 
diagram. In contrast 
to the leading-order perturbative mechanism, the latter, shown 
on the right panel of Fig.~\ref{fig0}, does not proceed through 
one-gluon exchange.
Although significant NLO corrections have been calculated~\cite{Belitsky2001, Diehl2007}, 
the higher-twist corrections dominate for the virtuality of the 
exchanged photon of the present data.

Alternatively, exclusive processes can be described at the hadronic level 
within the Regge formalism (e.g.~\cite{Reggereview}) where the interaction between 
the virtual photon and the nucleon is described in terms of Regge-trajectory 
exchanges in the $t$-channel. This approach can 
describe photoproduction~\cite{VGLphoto} 
and electroproduction~\cite{Laget} of pseudoscalar mesons 
 above the resonance region.
In the latter case, meson production by both transverse and longitudinal 
photons contribute and can be calculated within the Regge formalism.
 
This letter reports a measurement of the virtual-photon cross section for the hard exclusive 
reaction \reac on a hydrogen target. The relevant kinematic variables of 
the process are the squared four-momentum  $-Q^2$ of the exchanged 
virtual photon, either the Bjorken variable $x_B=Q^2/2M_p\nu$ (where $M_p$ is the proton mass and $\nu$ is the energy of the virtual photon in the target rest frame) or the squared invariant mass of the photon-nucleon system $W^2$,
the Mandelstam variable $t$, and the azimuthal angle $\phi$ of the pion around the virtual photon momentum relative to the lepton scattering plane.
Instead of $t$, the quantity $t^\prime=t-t_0$ was used in the analysis, where $-t_0$ represents the minimum value of $-t$ for a given value of $Q^2$ and $x_B$. 

The virtual-photon cross section has been previously measured above the resonance region (for $W^2>4$ GeV$^2$) at \cea\cite{CEA}, \cornell\cite{Cornell}, \desy\cite{Desy}, and more recently at \jlab\cite{HallC1,HallC2}. The present measurement extends the kinematic region to higher values of $W$ and higher values of $-t^\prime$.

The data were collected with the \hermes spectrometer~\cite{hermes:spectr} 
during the period 1996-2005. The 27.6 GeV \hera electron or positron
 beam at \desy 
scattered off polarized and unpolarized proton targets. Events were selected 
in which only one lepton and one positively charged hadron were detected and in which 
no additional cluster was recorded by the electromagnetic calorimeter. The 
\hermes geometrical acceptance 
of $\pm 170$ mrad horizontally and $\pm(40-140)$ mrad vertically results 
in detected scattering angles ranging from $40$ to $220$ mrad. Leptons 
were distinguished from hadrons with an average efficiency of $98\%$ and a
hadron contamination less than $1\%$ by using an electromagnetic calorimeter, a
transition-radiation detector, a preshower scintillation counter, and a
threshold gas \v{C}erenkov counter. In 1998 the threshold gas \v{C}erenkov counter was 
replaced by a Ring Imaging \v{C}erenkov detector~\cite{hermes:rich}. 
The threshold gas \v{C}erenkov counter (Ring Imaging \v{C}erenkov detector) provided
pion identification in the momentum range $4.9$ GeV $<p<15$ GeV ($1$ GeV $<p<15$ GeV). 
For the exclusive data sample, the pion momentum was required to be $7$ GeV $<p<15$ GeV. 
For this momentum range,
the pion identification efficiency is on average $97\%$ ($99\%$) and the contamination from 
other hadrons less than $3\%$ ($2\%$) for the threshold gas \v{C}erenkov 
counter (Ring Imaging \v{C}erenkov detector).	

The kinematic requirement $Q^2 > 1$~GeV$^2$ was imposed on the scattered 
lepton in order to select the hard scattering regime. 
The value of $W^2$ was required to be higher than $10$ GeV$^2$ to avoid the low-acceptance region for the hadron defined by the spectrometer upper angular limit of 220 mrad. 
The resulting kinematic range is $1$ GeV$^2<Q^2<11$ GeV$^2$ and $0.02<x_B<0.55$. The mean $W^2$ value of the data is 16 GeV$^2$.

\begin{figure}[t]
\begin{center}
\includegraphics[width=0.45\textwidth]{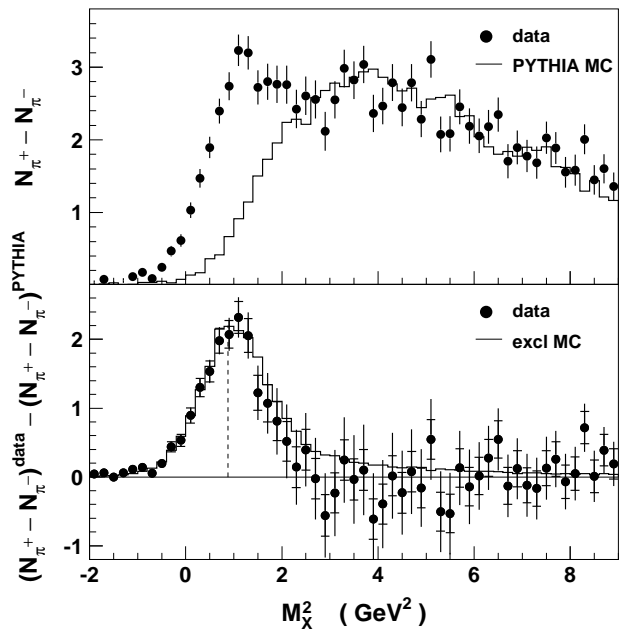}
\caption[]{
Upper panel: squared missing-mass dependence of the normalized-yield difference
$\sf{N}_{\pi^+}-\sf{N}_{\pi^-}$ for data (filled
points) and \pythia Monte Carlo (histogram). The error bars represent the statistical uncertainty. Lower panel: squared missing-mass
dependence of the normalized-yield after background subtraction procedure. The data (filled points) are compared to a Monte Carlo sample for exclusive $\pi^+$ production (histogram) normalized 
to the data. The inner
error bars represent the statistical uncertainties and the outer error bars
represent the quadratic sum of statistical and systematic uncertainties. 
The latter originate from the background subtraction procedure. 
The dashed vertical line indicates the squared neutron mass.}
\label{fig1}
\end{center}
\end{figure}

As the recoiling neutron was not detected, exclusive meson production 
was selected by requiring that the squared missing mass $M^2_X$ 
of the reaction \reacX corresponds to the squared neutron mass. 
Due to the limited experimental resolution, the exclusive $\pi^+$ channel 
cannot be separated from the neighboring channels (defined as background
channels) with final states such as $\pi^+ + (N\pi)$ and $\pi^+ + (N\pi\pi)$, 
as their $M^2_X$ values can be smeared into the region corresponding to exclusive $\pi^+$ channel. 
In order to subtract the background channels, which cannot be fully described by existing Monte Carlo simulations, a two-step procedure was developed.
In the first step, the yield difference between $\pi^+$ and $\pi^-$ was used.
In this yield difference, exclusive $\pi^+$ events remain as the exclusive
production of $\pi^-$ on a hydrogen target with a recoiling nucleon in the
final state is forbidden by charge conservation. Background events arising from
the production of neutral vector mesons cancel as they contribute equally
to $\pi^+$ and $\pi^-$ production.  
In the second step, the \pythia Monte Carlo generator~\cite{PYTHIA6} was used 
in conjunction with a special set of \jetset\cite{ph:jetset} fragmentation 
parameters tuned to provide an accurate description of 
deep inelastic hadron production in the \hermes kinematic domain.
The difference $\sf{N}_{\pi^+}-\sf{N}_{\pi^-}$, where $\sf{N}$ represents the yield normalized by the integrated luminosity $\mathcal{L}$, is well described by \pythia if pion momenta larger than 7 GeV are required. The latter constraint removes mainly background events for which $Q^2<3$ GeV$^2$. The upper panel of Fig.~\ref{fig1} shows the $M_X^2$ dependence of the normalized-yield difference $\sf{N}_{\pi^+}-\sf{N}_{\pi^-}$ for the data and for the \pythia Monte Carlo simulation. The Monte Carlo sample describes the data for values of the squared missing mass higher than 2 GeV$^2$, i.e., outside the region corresponding to exclusive $\pi^+$ production. 
Finally, the exclusive $\pi^+$ yield was obtained by subtracting the 
normalized-yield difference of
the \pythia Monte Carlo from that of the data: $\sf{N}^{excl}_{\pi^+}=(\sf{N}_{\pi^+}-\sf{N}_{\pi^-})^{data}-(\sf{N}_{\pi^+}-\sf{N}_{\pi^-})^{PYTHIA}$. 
The lower panel of Fig.~\ref{fig1} shows the peak corresponding to exclusive $\pi^+$ production resulting from
this double difference. The peak is centered at the squared neutron mass and its
width of 0.67 GeV$^2$ is consistent with that of a Monte Carlo sample for exclusive $\pi^+$ production normalized to the data. The Monte Carlo, denoted as exclusive Monte Carlo, is based on a GPD model~\cite{VGG} and will be described below. 
As \pythia does not simulate nucleon resonance production, 
the agreement of the
exclusive Monte Carlo sample with the data for the double difference is an
indication that there is very little contribution from resonant channels ($\pi^+
+ \Delta^0$ for $\pi^+$ and $\pi^- + \Delta^{++}$ for $\pi^-$) to the normalized-yield difference $\sf{N}_{\pi^+}-\sf{N}_{\pi^-}$.
In order to estimate the systematic uncertainty of the background subtraction, the \pythia Monte Carlo normalized-yield difference $(\sf{N}_{\pi^+}-\sf{N}_{\pi^-})^{PYTHIA}$ was changed by the discrepancy between
that normalized-yield and the data in the region $3<M^2_X<7$ GeV$^2$. 
The discrepancy amounts to between $20\%$ and $50\%$,  
depending on the specific kinematic bin in $Q^2$, $x_B$, or $t^\prime$.
The largest discrepancies correspond to the least populated bins ($Q^2>4$ GeV$^2$ and $-t^\prime>0.3$ GeV$^2$). 
An upper limit on $M_X^2$ of 1.2 GeV$^2$ was chosen in order to optimize
the combined statistical and systematic uncertainties for the number of
exclusive events. The resulting relative systematic uncertainty for the number
of exclusive events ranges from $5\%$ to $20\%$. Within the $M_X^2$ limit, the
fraction of background events in the normalized-yield difference
$\sf{N}_{\pi^+}-\sf{N}_{\pi^-}$ estimated with \pythia amounts to $20\%$. The
number of $\pi^+$ events after background subtraction is 4510. 

As the recoiling
neutron was not detected, $t^\prime$ was determined from the measurement of
the four-momenta of the scattered lepton and the produced pion. In order to improve the resolution for exclusive events,
$t^\prime$ was calculated by setting $M_X=M_n$. This offered the possibility 
to discard the lepton energy, the quantity subject to the largest uncertainty 
for the present data sample. 
This procedure results in a factor of two improvement in the $t^\prime$
resolution. 
The same
calculation was applied to the \pythia Monte Carlo events that were used to 
subtract the background. 

The differential cross section for exclusive $\pi^+$ production by virtual photons can be written as 
\begin{eqnarray}
\nonumber
\frac{d\sigma^{\gamma^* p\to n \pi^+ }(x_B,Q^2,t^\prime,\phi)}{dt^\prime d\phi}= \\
\frac{1}{\Gamma_V(x_B,Q^2)}\frac{d\sigma^{e p\to e n \pi^+ }(x_B,Q^2,t^\prime,\phi)}{dx_BdQ^2dt^\prime d\phi},
\end{eqnarray}
\noindent
where $\Gamma_V$ is the virtual-photon flux factor. 
Within the Hand convention~\cite{Hand}, this flux factor is
\begin{eqnarray}
\Gamma_V(x_B,Q^2)=\frac{\alpha}{8\pi}\frac{1}{M_p^2E^2}\frac{Q^2}{x_B^3}\frac{1-x_B}{1-\epsilon},
\end{eqnarray}
\noindent
where $E$ is the beam energy and $\epsilon$ is the virtual-photon polarization parameter. 
The $t^\prime$ dependence of the photon-nucleon differential cross section integrated over
$\phi$ were extracted from the data as follows:
\begin{eqnarray}
\frac{d\sigma^{\gamma^* p\to n \pi^+ }(x_B,Q^2,t^\prime)}{dt^\prime}=\frac{1}{\Gamma_V}.\frac{N^{excl}_{\pi^+}}{\mathcal{L} ~\Delta x_B ~\Delta Q^2 ~\Delta t^\prime ~\kappa~\eta},
\label{eq1}
\end{eqnarray}
\noindent
where $N^{excl}_{\pi^+}$ is the number of $\pi^+$ events after background
subtraction, $\kappa$ is the 
probability to detect the scattered lepton and the produced 
$\pi^+$ within the \hermes spectrometer acceptance, 
and $\eta$ is the radiative correction factor. These quantities were determined for each kinematic bin. 
The symbols $\Delta x_B$, $\Delta Q^2$, and $\Delta t^\prime$ denote the 
bin size, which was chosen according to the statistical precision, instrumental
resolution, and kinematic smearing affecting each individual bin.
The $t^\prime$ dependence of the differential cross section $\frac{d\sigma(x_B,Q^2,t^\prime)}{dt^\prime}$ was determined for four $Q^2$ bins and the $Q^2$ dependence of the cross section integrated over $t^\prime$, $\sigma(x_B,Q^2)$, for three $x_B$ bins. 

\begin{figure*}[ht]
\begin{center}
\includegraphics[width=0.95\textwidth]{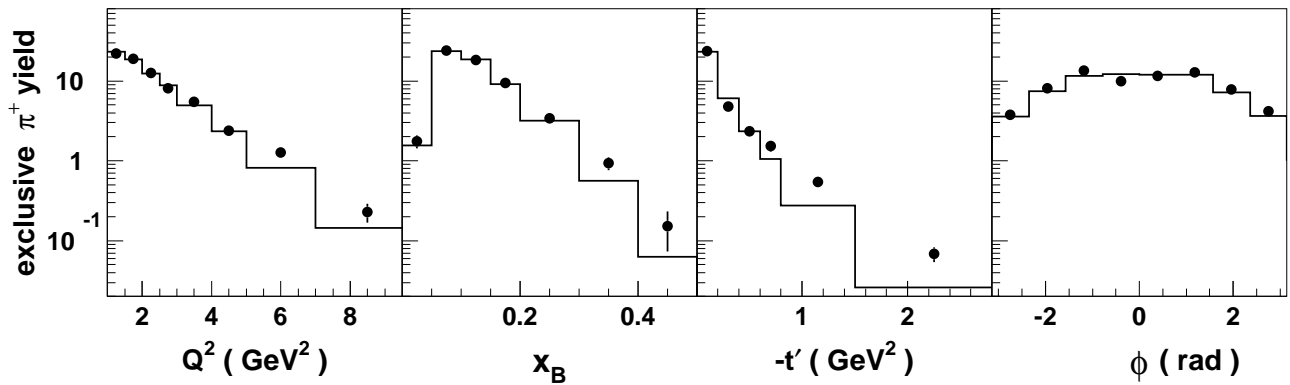}
\caption[]{
Distributions of exclusive $\pi^+$ events within the \hermes acceptance 
as a function of $Q^2$, 
$x_B$, $-t^\prime$ and $\phi$ for data (filled points) compared with an exclusive 
Monte Carlo simulation (histogram) based on a GPD parametrization~\cite{VGG} 
which was tuned to the data (see text).
The data sample is arbitrarily normalized; the Monte Carlo sample is 
normalized to the data using a constant factor. 
The error bars represent the statistical and systematic uncertainties 
of the background subtraction added in quadrature.}
\label{fig2}
\end{center}
\end{figure*}

The detection probability $\kappa$ was determined using an exclusive $\pi^+$
Monte Carlo simulation based on a GPD model~\cite{VGG} and a \geant
simulation~\cite{GEANT} of the detector. The variable $\kappa$ is the ratio between the
number of simulated events reconstructed in the \hermes spectrometer and the
number of generated events. These numbers were evaluated at the reconstructed
and generated kinematics respectively and were integrated over the bin size
and over $\phi$ (and $t^\prime$ for the determination of $\sigma(x_B,Q^2)$).
The variable $\kappa$ represents the combined effect of the spectrometer acceptance, the
analysis constraints (such as the $M^2_X$ constraint) and the  
detector efficiencies.
For the GPD model~\cite{VGG}, the power corrections were included and the factorized ansatz was used for the $t^\prime$ dependence. However, the $t^\prime$ and $\phi$ dependences were modified in order to describe the data. 
Radiative events were included in the Monte Carlo simulation according to the code \radgen\cite{hermes:radgen} adapted to exclusive meson production using the GPD model~\cite{VGG}. 
Fig.~\ref{fig2} shows good agreement between the kinematic distributions 
of the data and of the exclusive Monte Carlo sample (which is normalized to the data 
by a constant factor) with the exception of yields at high values 
of $Q^2$, $x$ and especially$-t^\prime$.

The calculated spectrometer acceptance ranges from 0.1 to 0.7 depending on 
the kinematic bin, while the efficiency of the detectors and 
analysis constraints together amounts to about 0.4--0.5. The combination of both 
leads to $\kappa$ values ranging from 0.04 to 0.28.
The model dependence of the determination of
$\kappa$ was estimated by using the GPD parametrization~\cite{VGG}
with different $t^\prime$ and  $\phi$  dependences and by using a
different GPD parametrization~\cite{Piller} for the $Q^2$ and $x_B$
dependences. It was further studied to what degree the relevant
kinematic variables $Q^2$, $t^\prime$ and $\phi$ are 
correlated by the detector acceptance.  These model dependences
amount to a relative systematic uncertainty in $\kappa$ of less than $15\%$.

Events can be smeared from one bin to another
by multiple scattering and bremsstrahlung. This
effect was included in the determination of 
the detection probability $\kappa$. 
With the selected bin sizes, the
fraction of events that migrate out of (into) a certain kinematic bin is on
average $12\%$ ($15\%$) and is always below $25\%$ ($35\%$) according to the 
exclusive Monte Carlo simulation.

The Born cross section was extracted using 
the radiative correction factor 
$\eta$, which was determined from the ratio of Monte Carlo samples with and 
without radiative effects. 
The value of $\eta$ was found to be $0.77$ with little variation (less than $3\%$) 
as a function of $Q^2$, $x_B$ or $t^\prime$ with the constraints 
applied to the data to select exclusive $\pi^+$ production.

The mean values for $x_B$, $Q^2$, and $t^\prime$ for each
kinematic bin were estimated from distributions generated at Born level
by the exclusive Monte Carlo simulation. 
The flux factor in Eq.~\ref{eq1} was determined for these mean values 
of $Q^2$ and $x_B$. 
The resulting cross sections were corrected for bin-averaging effects to take into
account the nonlinear dependence of the cross section within each bin
(in Eq.~\ref{eq1},
a linear dependence inside a bin is implicitly assumed). These corrections were
obtained from the exclusive Monte Carlo simulation by taking the ratio 
between the cross section evaluated at fixed kinematic values and 
the cross section integrated
over the kinematic bins following Eq.~\ref{eq1}. The error arising from 
the evaluation of the flux factor at the mean values is then also corrected.
The bin-averaging correction factor amounts on average to $1.08$ and does not exceed $1.2$
except for the highest-$Q^2$ bin, where it reaches $2.2$ at some
$t^\prime$ values.

The integrated luminosity $\mathcal{L}$ was determined by comparing the number 
of inclusive deep-inelastic scattering events in the data sample to the yield 
generated by a Monte Carlo sample for inclusive scattering 
based on world data~\cite{F2ALLM1,F2ALLM2}. 
The \hermes luminosity detector provided another measurement of $\mathcal{L}$, which 
was used to estimate the systematic uncertainty. The integrated luminosity $\mathcal{L}$ amounts to 
0.4~fb$^{-1}$ with a 5$\%$ systematic uncertainty.

The total systematic uncertainty of the cross section 
is dominated by the uncertainty of the 
background subtraction and of the detection probability. 
The latter takes into account the uncertainties due to the model dependence 
 of its determination and the different detector resolutions in the different 
data taking periods.

The measured differential cross section integrated over the angle $\phi$ can be written 
as $\frac{d\sigma}{dt} = \frac{d\sigma_T}{dt}\ + \epsilon \frac{d\sigma_L}{dt}$ 
where $\sigma_T$ and $\sigma_L$ are, respectively, the contributions of 
transversely and longitudinally polarized virtual photons. 
At \hermes the separation of the transverse and longitudinal components of 
the cross section is not feasible. 
However, as the transverse contribution is predicted to be suppressed by
 $1/Q^2$ with respect to the longitudinal contribution~\cite{Collins}, 
the data at larger $Q^2$ are expected to be
dominated by the longitudinal part. 
Moreover, with the presence of the pion pole 
at low $-t^\prime$, the longitudinal part of the cross section is expected 
to dominate in this region. 
In Fig. 4 and 5, described below,
the data are compared to calculations for the longitudinal part of the cross
section computed using a GPD model~\cite{VGG} and to calculations for the
total cross section computed using a Regge model~\cite{Laget2}. 
The cross sections are calculated for the mean values of $x_B$, $Q^2$, and $t^\prime$ 
of the experimental data in each bin.
\begin{figure*}[ht]
\begin{center}
\includegraphics[width=0.95\textwidth]{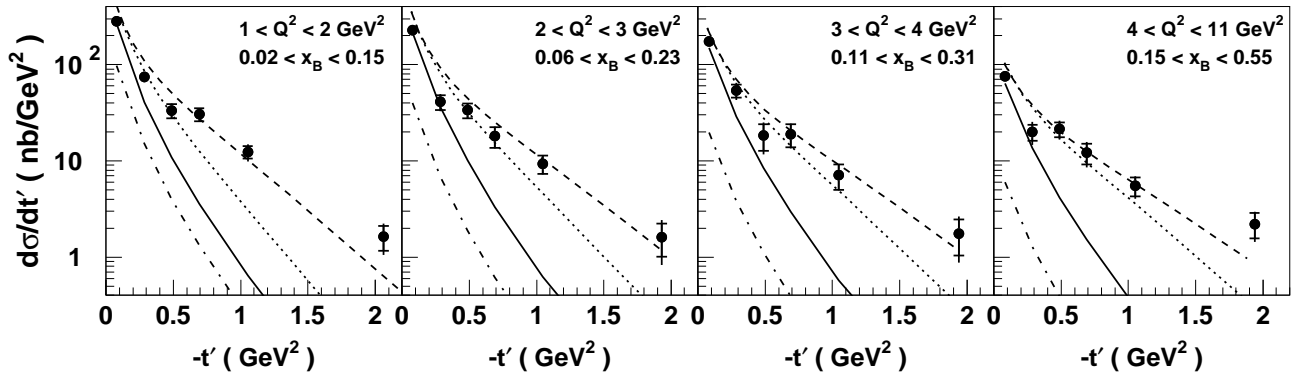}
\caption[]{Differential cross section for exclusive $\pi^+$ production by virtual photons 
as a function of $-t^\prime$ for four $Q^2$ bins. 
The inner error bars 
represent the statistical uncertainties and the outer error bars represent
the quadratic 
sum of statistical and systematic uncertainties. The curves represent 
calculations based on a GPD model~\cite{VGG} for $\frac{d\sigma_L}{dt^\prime}$ using a 
Regge-type ansatz for the $t^\prime$ dependence 
(dashed-dotted lines: leading-order calculations, solid lines: with power corrections) 
and a Regge model~\cite{Laget2} for $\frac{d\sigma}{dt}$ (dashed lines) and $\frac{d\sigma_L}{dt}$ (dotted lines).}
\label{fig3}
\end{center}
\end{figure*}

\begin{figure}[t]
\begin{center}
\includegraphics[width=0.45\textwidth]{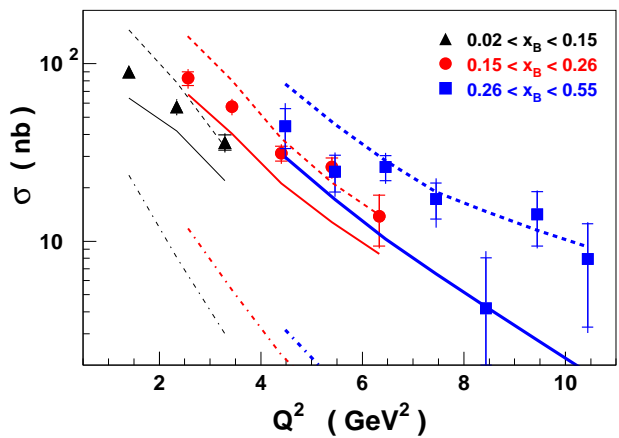}
\caption[]{Cross section for exclusive $\pi^+$ production by virtual photons 
as a function of $Q^2$ for three $x_B$ bins and integrated over $t^\prime$ (colour online). 
The inner error bars 
represent the statistical uncertainties and the outer error bars represent
the quadratic 
sum of statistical and systematic uncertainties. The curves represent 
calculations based on a GPD model~\cite{VGG} for $\sigma_L$ using a 
Regge-type ansatz for the $t^\prime$ dependence (dashed-dotted lines: leading-order 
calculations, solid lines: with power corrections) and 
a Regge model~\cite{Laget2} for $\sigma$ (dashed lines). The thin, medium and thick 
lines correspond to the low, medium and high $x_B$ values, respectively.}
\label{fig4}
\end{center}
\end{figure}
Fig.~\ref{fig3} shows the $t^\prime$ dependence of the differential cross section 
for four $Q^2$ bins. As $Q^2$ is closely related to $x_B$ 
(due to the \hermes
acceptance as well as the upper limit on the pion momentum)
low values of $Q^2$ correspond to low values of $x_B$. 
The dashed-dotted lines in Fig.~\ref{fig3} show the leading-order 
calculations of the
longitudinal part computed using the GPD model~\cite{VGG}. 
The GPD $\widetilde E$ is considered to be dominated by the $t-$channel 
pion-pole and the pseudoscalar contribution to the cross 
section is parametrized in terms 
of the pion electromagnetic form factor $F_\pi$.
 A Regge-inspired
$t^\prime$ dependence is used for $\widetilde E$. 
The GPD $\widetilde H$ is neglected here as $\widetilde E$ is
expected to dominate the cross section at low $-t^\prime$. 
The solid lines include the power corrections due to the intrinsic transverse 
momentum of the partons and due to the
soft-overlap contribution, the latter being dominant.
While the leading-order calculation strongly underestimates the data, 
the calculations including 
power corrections agree with the data for $-t^\prime<0.3$ GeV$^2$ for the four $Q^2$ bins. 
As the GPD model requires $-t^\prime$ to 
be much smaller than $Q^2$, 
the calculations are not expected to describe the full $t^\prime$ range. 
Furthermore, at larger $-t^\prime$, the data may receive a significant 
contribution from the transverse part of the cross section, which is not described 
by the GPD model.
Fig.~\ref{fig4} shows the $Q^2$ dependence of the cross section integrated over $t^\prime$ for three $x_B$ bins. 
The $Q^2$ dependence of the data is in general well described by the calculations from the 
GPD model~\cite{VGG} 
with the inclusion of the power corrections, although the magnitude of the cross section 
is underestimated. 
The data support the order of magnitude of the power corrections for the 
calculations by the GPD model~\cite{VGG} at 
low $-t^\prime$, a region where the longitudinal part of the cross section is 
expected to dominate, and for the available $Q^2$ range. 

Both the transverse and longitudinal parts of the cross section were computed 
using a Regge model~\cite{Laget2}, where pion production is described by the 
exchange of $\pi$ and $\rho$ Regge trajectories. In this formalism, 
the meson-nucleon coupling constants are fixed by pion photoproduction data. 
In the original version~\cite{Laget}, the $\pi\pi\gamma$ form 
factor is fixed by pion form factor measurements, while the $\pi\rho\gamma$ 
transition form factor is unconstrained. In the version used 
here~\cite{Laget2}, both form factors are taken to be both $Q^2$-
and $t^\prime$-dependent. 
The dashed (dotted) lines in Fig.~\ref{fig3} show the total 
(longitudinal) cross section computed using the Regge model. In this model the 
transverse part of the cross section is estimated to represent 
from $6\%$ to $8\%$ of the total cross section at 
$-t^\prime=0.07$ GeV$^2$ and from $15\%$ to $25\%$ 
of the total cross section integrated over $t^\prime$, confirming 
the expected suppression of the transverse to the longitudinal part 
of the cross section. 
However data from \jlab\cite{HallC2} at lower center of mass energy 
($W^2=4.9$ GeV$^2$) show that the transverse part of the 
cross section is underestimated by the Regge model by a factor 
of three to four. It is not clear if this also holds at the
higher $W^2$ and $-t^\prime$ values of the \hermes data.
Compared to the calculations for the longitudinal cross section 
from the Regge model (dotted lines), the $t^\prime$ dependence 
of the GPD model~\cite{VGG} (solid lines) in Fig.~\ref{fig3} appears too steep. 
The total cross section computed by the Regge model describes well the $t^\prime$ 
dependence of the differential cross section (dashed line on Fig.~\ref{fig3}) 
and the $Q^2$ dependence of the cross section integrated over $t^\prime$ (dashed lines on Fig.~\ref{fig4}).
The model calculations give also a good description of the magnitude 
of the data except at low $-t^\prime$ for $Q^2<3$ GeV$^2$, 
where the calculations overestimate the data up to $70\%$.

In conclusion, the cross section was measured for exclusive
electroproduction of $\pi^+$ mesons from a hydrogen target as a function of
$-t^\prime$ for four $Q^2$ bins and as a function of $Q^2$ for three $x_B$ bins. 
A model calculation for the longitudinal part of the virtual-photon cross 
section based on the Generalized Parton Distributions~\cite{VGG} 
does not describe the data at the leading order in the present $Q^2$ range. 
However, if power corrections are included, the model is in fair agreement 
with the magnitude of the data at low values of $-t^\prime$. 
In this kinematic region, where $\pi^+$ production by longitudinal photons
is expected to dominate, 
the data support the order of magnitude of the power corrections. 
A model calculation based on the Regge formalism for both the longitudinal 
and the transverse part of the cross section~\cite{Laget2} 
provides a good description of the magnitude and the $t^\prime$ and the $Q^2$ 
dependences of the data.

\begin{acknowledgments} 
 
We thank M. Guidal, J.M. Laget, and M. Vanderhaeghen for the theoretical calculations 
as well as M. Diehl 
for many interesting and useful discussions.
We gratefully acknowledge the \desy management for its support, the staff
at \desy and the collaborating institutions for their significant effort,
and our national funding agencies and the EU RII3-CT-2004-506078 program
for financial support.
\end{acknowledgments}

\end{document}